\documentclass[final,3p,times]{elsarticle}
\usepackage{amsmath}
\usepackage{graphicx}
\usepackage{algorithmic}
\usepackage{algorithm}
\usepackage{amssymb}
\usepackage[tight,footnotesize]{subfigure}
\usepackage{array}
\usepackage{color}
\usepackage{multirow} \usepackage{rotating}

\DeclareGraphicsExtensions{.eps}

\newcolumntype{C}[1]{>{\raggedright\let\newline\\\arraybackslash}p{#1}}

\def\x{{\mathbf x}}
\def\v{{\mathbf v}}

\def\z{{\mathbf z}}

\def\RP{{\mathcal{A}}}
\def\M {{\mathbf{M}}}         
\def\OM {{\mathbf{\Omega}}}    
\def\x {{\mathbf{x}}}        
\def\e {{\mathbf{e}}}        
\def\y {{\mathbf{y}}}        
\def\w {{\mathbf{w}}}        
\def\mdim {m}   
\def\sdim {d}   
\def\cosp {\ell} 
\def \D {\mathbf{D}}         
\def \I {\mathbf{I}}         

\newtheorem{thm}{Theorem}[section]
\newtheorem{cor}[thm]{Corollary}

\newtheorem{defn}[thm]{Definition}
\newtheorem{rem}[thm]{Remark}

\newcommand{\norm}[1]{\left\Vert#1\right\Vert}

\newcommand\vect[1]{{\bf#1}}
\newcommand\matr[1]{{\bf#1}}

\newcommand\alphabf{{\boldsymbol{\alpha}}}
\newcommand{\argmin}{\operatornamewithlimits{argmin}}

\newcommand\RR[1]{\mathbb{R}^{#1}}

\DeclareMathOperator{\rank}{rank}

\newcommand{\rg}[1]{\textcolor{black}{#1}}

\journal{Draft}

\begin{document}

\begin{frontmatter}

\title{Sampling in the Analysis Transform Domain }


\author[csa]{Raja~Giryes}
\ead{raja.giryes@duke.edu.}

\address[csa]{The Department of Electrical and Computer Engineering,
        Duke University,
        Durham, NC 27708, USA}

%




\begin{abstract}

Many signal and image processing applications have benefited remarkably from the fact that the underlying signals reside in a low dimensional subspace. One of the main models for such a low dimensionality is the sparsity one. Within this framework there are two main options for the sparse modeling: the synthesis and the analysis ones, where the first is considered the standard paradigm for which much more research has been dedicated. In it the signals are assumed to have a sparse representation under a given dictionary. On the other hand, in the analysis approach the sparsity is measured in the coefficients of the signal after applying a certain transformation, the analysis dictionary, on it. Though several algorithms with some theory have been developed for this framework, they are outnumbered by the ones proposed for the synthesis methodology. 

Given that the analysis dictionary is either a frame or the two dimensional finite difference operator, we propose a new sampling scheme for signals from the analysis model that allows recovering them from their samples
using any existing algorithm from the synthesis model.
The advantage of this new sampling strategy is that it makes the existing synthesis methods with their theory also available for signals from the analysis framework.
\end{abstract}


\end{frontmatter}



\section{Introduction}
\label{sec:intro}

The idea that signals reside in a union of low dimensional subspaces has been used extensively in the recent decade in many fields and applications \cite{Bruckstein09From}. One of the main problems that has benefited remarkably from this theory is the one of compressed sensing. In this problem we want to recover an unknown signal $\x \in \RR{d}$ from a small number of noisy linear measurements:
\begin{eqnarray}
\label{eq:lin_meas}
\y = \M \x + \e,
\end{eqnarray}
where $\M \in \RR{m \times d}$ is the measurements matrix, \rg{$\e \in \RR{m}$} is an additive noise and $\y \in \RR{m}$ is the noisy measurement.     

If the signal $\x$ can be any signal then we are in a hopeless situation in the task of recovering it from $\y$. However, if we restrict it to a low-dimensional manifold that does not intersect with the null space of $\matr{M}$ at any point except the origin then we are more likely to be able to recover $\x$ from $\y$ by looking for the signal at this manifold, which is closest to $\y$ after multiplying it by $\M$. 

An example for such a low dimensional manifold is the one of $k$-sparse signals under a given dictionary $\D \in \RR{d \times n}$. In this case our signal $\x$ satisfies 
\begin{eqnarray}
\x = \D \alphabf, \norm{\alphabf}_0 \le k,
\end{eqnarray} 
where $\norm{\alphabf}_0$ is the $\ell_0$-pseudo norm that counts the number of non-zero entries in a vector. In this case we may recover $\x$ from $\y$ by minimizing the following problem,
\begin{eqnarray}
\label{eq:l0_synthesis}
\hat\alphabf_{S-\ell_0} = \argmin_{\tilde\alphabf}\norm{\tilde\alphabf}_0 & s.t &
\norm{\y - \M\D\tilde{\alphabf}}_2 \le \lambda_{\e}. 
\end{eqnarray}
where $\lambda_{\e}$ is an upper bound for  $\norm{\e}_2$ if the noise is bounded and adversarial, or a scalar dependent on the noise distribution \cite{Candes06Modern}.
As this problem is NP-hard \cite{Davis97Adaptive} many approximation methods have been proposed for it \cite{Elad10Sparse, Foucart13Mathematical}, such as orthogonal matching pursuit (OMP) \cite{MallatZhang93} and the $\ell_1$-relaxation strategy that replaces the $\ell_0$-pseudo norm with the $\ell_1$-norm in \eqref{eq:l0_synthesis} \cite{Candes05Decoding}.

One of the main theoretical questions being asked with regard to these algorithms is what are the requirements on $\M$, $\D$, $m$ and $k$ such that the representation, $\alphabf$, of  $\x$ may be stably recovered from $\y$ using these techniques, i.e., their recovery $\hat{\alphabf}$ will satisfy 
\begin{eqnarray}
\label{eq:stab_rec}
\norm{\hat{\alphabf} - \alphabf}_2 \le C \norm{\vect{e}}_2,
\end{eqnarray} 
where $C$ is a certain constant (different for each algorithm).
  
Two main tools have been used to answer this question. The first is the coherence of $\M\D$ \cite{Donoho06Stable}, which is the maximal (normalized) inner product between the columns of $\M\D$. It has been shown that if the matrix $\M\D$ is incoherent (has a small coherence) then it is possible to get a stable recovery using OMP and the $\ell_1$-relaxation. The problem with the coherence based recovery conditions is that they limit the number of measurements $m$ to be of the order of $k^2$, while $m = 2k$ is enough to guarantee uniqueness for \eqref{eq:lin_meas} in the noiseless case and $m = O(k \log(n))$ is enough for stability in the noisy one. 

The second property of $\M\D$ used to derive reconstruction performance guarantees is the restricted isometry property (RIP). This property provides us with a bound on the minimal and maximal eigenvalues of every sub-matrix consisting of any $k$-columns from a given matrix. Formally,
\begin{defn}[RIP \cite{Candes06Near}]
\label{def:RIP}
A matrix $\matr{A} \in \RR{m \times n}$ has the RIP with a constant $\delta_{k}$,
if $\delta_{k}$ is the smallest constant that satisfies
\begin{eqnarray}
\label{eq:D_RIP}
&& (1-\delta_{k})\norm{\tilde\alphabf}_2^2 \le \norm{\matr{A}\tilde\alphabf}_2^2 \le (1+\delta_k)\norm{\tilde\alphabf}_2^2,
\end{eqnarray}
whenever $\tilde\alphabf \in \RR{n}$ is $k$-sparse.
\end{defn}
It has been shown for many approximation algorithms that they get stable recovery in the form of \eqref{eq:stab_rec}, if $\M\D$ has the RIP with a constant $\delta_{ak} < \delta_{ref}$, where $a$ and $\delta_{ref}$ are two constants dependent on the algorithm in question  \cite{Candes06Near, Needell09CoSaMP, Dai09Subspace, Blumensath09Iterative, Foucart11Hard, Zhang11Sparse}. 
The true force behind these RIP conditions is that it has been shown that many matrices (typically random subgaussian matrices) satisfy this bound given that $m = O(k \log(n))$ \cite{Candes06Near, Rudelson06Sparse, Baraniuk08Simple}. 
Notice that the main significance of this result is that it shows that it is possible to recover a signal from a number of measurements proportional to its manifold dimension $k$.

An alternative model for low dimensional signals that relies on sparsity is the analysis  framework \cite{elad07Analysis, Nam12Cosparse}. In this paradigm, we look at the behavior of the signal after applying a certain operator $\OM \in \RR{n \times d}$ on it, assuming that $\OM\x$ has $\cosp$ zeros. \rg{The number of zeros, $\cosp$, is termed the cosparsity of the signal $\x$ \cite{Nam12Cosparse}}. 
With this prior at hand, we may recover $\x$ from \eqref{eq:lin_meas} by solving 
\begin{eqnarray}
\label{eq:l0_analysis}
\hat\x_{A-\ell_0} = \argmin_{\tilde\x}\norm{\OM\tilde\x}_0 & s.t &
\norm{\y - \M\tilde{\x}}_2 \le \lambda_{\e},
\end{eqnarray}
where here also $\lambda_\e$ depends on the noise properties. 

Note that as we minimize the number of non-zeros in $\OM\x$ in \eqref{eq:l0_analysis}, the number of zeros is the one that defines the manifold dimension in which $\x$ resides. Each zero in $\OM \x$ corresponds to a row in $\OM$ to which $\x$ is orthogonal. 
Denoting by $T$ the support of $\OM\x$ and $T^C$ it complimentary, we may say that $\x$ resides in a subspace of dimension $d - \rank(\OM_{T^C})$. Therefore if $\OM\x$ has $\cosp = n- k $ zeros, where $k$ is the number of non-zeros in it, and $\OM$ is in general position, i.e., every $d$ rows in it are independent, then the manifold dimension is $d-\cosp$.

In the noiseless case ($\e = 0$), the requirement $ m = 2 \left(d - \rank\left(\OM_{T^C} \right)\right)$ is enough to guarantee uniqueness in the solution of \eqref{eq:l0_analysis} (and therefore the recovery of $\x$) under very mild assumptions on the relation between $\OM$ and $\M$ \cite{Nam12Cosparse}. However, in the noisy case having a number of samples at the order of the manifold dimension, i.e., $m = O\left(d - \rank\left(\OM_{T^C} \right)\right)$  is not enough to guarantee stability even by solving \eqref{eq:l0_analysis} \cite{Giryes14effective}.
Therefore, it is not surprising that the recovery conditions for algorithms that approximate \eqref{eq:l0_analysis} require $m = O( k \log(n) )$ \cite{Candes11Compressed, Liu12Compressed, Giryes14Greedy, Needell13Stable, Needell13Near, Kabanava13Analysis, Giryes14GreedyAlgorithm}, where $\OM$ is assumed to be either a frame \cite{Candes11Compressed, Liu12Compressed, Kabanava13Analysis, Giryes14GreedyAlgorithm}, the 2D-DIF operator \rg{\cite{Needell13Stable, Needell13Near, Krahmer14Stable, Poon14TV} } or an operator that  generates a manifold with a tractable projection onto it \cite{Giryes14Greedy}.

Though the number of measurements in synthesis and analysis are similar there are two major differences between the two: (i) In synthesis the number of measurements are proportional to the manifold dimension, while in analysis this is not necessarily the case as $k = n -\cosp$ might be remarkably larger than $ d - \rank( \OM_T^C )$ (See \cite{Giryes14Greedy} for more details); (ii) In synthesis the dictionary $\matr{D}$ must be incoherent as otherwise the RIP condition will no longer hold \cite{Rauhut08Compressed}, while in the analysis case there is no such restriction on the analysis dictionary $\OM$ but only on $\matr{M}$. 

An interesting relation between analysis and synthesis, which is depicted in \cite{elad07Analysis}, is that if $\OM$ is a frame and $\OM\x$ is $k$-sparse then $\x$ has a $k$-sparse representation under $\matr{D} = \OM^{\dag}$ (the pseudo-inverse of $\OM$), i.e., $\x = \matr{D}\OM\x$. Therefore, if $k$ is small enough then relying on the uniqueness of the sparse representation \cite{Donoho02Optimally}, we can recover $\x$ by minimizing \eqref{eq:l0_synthesis}. The problem we encounter in this case is that unless $\OM$ is an incoherent matrix (its rows are incoherent) and therefore $\matr{D}$ is incoherent, \rg{none} of the existing synthesis approximation algorithms is guaranteed to provide us with a good estimate\footnote{Some recent works have addressed the case of coherent dictionaries in the synthesis case \cite{Davenport13Signal, Giryes14GreedySignal, Giryes13CanP0, Giryes13IHTconf, Giryes13OMP, Hegde14Approximation}. However, they are very limited to specific cases and do not apply to general types of dictionaries such as frames.}. 

\subsection{Our Contribution}

In this work we provide a new sampling strategy that allows recovering signals from the analysis model using any existing synthesis algorithm, given that the analysis dictionary is either a frame or the 2D-DIF operator. 
Our scheme is general and can be easily extended to other types of analysis dictionaries. Instead of sampling the signal itself, we sample the signal in the analysis transform domain and then perform the recovery in this domain. From the proxy in the transform domain we get a reconstruction of our original signal. 
The idea to recover an analysis signal in the transform domain is not a new idea and was used before  \cite{Giryes14GreedyAlgorithm, Ophir11Multi,Ravishankar11MR,Ravishankar13Sparsifying}. However, the uniqueness in our approach compared to previous works is that (i)
we sample with one matrix and then use another one for recovery; and (ii) we make use of existing synthesis algorithms as a black box without changing them for recovering the transform domain coefficients of the signal.
Our sampling and recovery strategy is presented in Section~\ref{sec:sampling_scheme} for the case that the analysis dictionary is a general frame or the 2D-DIF operator. 
In Section~\ref{sec:exp} we provide a simple demonstration of the usage of our scheme and in Section~\ref{sec:conc} we conclude the paper. 

\section{Sampling in the Transform Domain}
\label{sec:sampling_scheme}

Before we turn to present our scheme let us recall the problem we aim at solving in the analysis case:
\begin{defn}[Problem $\cal{P}$]
Consider a measurement vector $\y \in \RR{\mdim}$
such that $\y=\M\x + \e$ where \rg{$\OM\x\in \RR{n}$} is either $k$-sparse
for a given and fixed analysis operator $\OM \in \RR{n \times \sdim}$
or almost $k$-sparse, i.e. $\OM \x$ has $k = n - \cosp$ leading elements.
The non-zero locations of the $k$ leading elements is denoted by $T$.
$\M\in \RR{\mdim\times \sdim}$ is a degradation operator and $\e\in \RR{\mdim}$ is an additive noise.
Our task is to recover $\x$ from $\y$. The recovery result is denoted by $\hat\x$.
\end{defn}

\subsection{Guarantees for Frames}

Let $\matr{A} \in \RR{m \times n}$ be a given matrix and $\RP(\y) = \RP(\y| \matr{A},k)$ be an algorithm that receives a signal $\vect{y}$ such that $\vect{y} = \matr{A}\alphabf + \vect{e}$, where $\alphabf \in \RR{n}$ is either $k$-sparse or  almost $k$-sparse, such that either one of the following (or the two of them) holds: (i) for the case that ${\vect{e}}$ is an adversarial noise with a bounded energy 
it is guaranteed that
\begin{eqnarray}
\label{eq:synthesis_advers_noise_bound}
\norm{\alphabf - \RP(\y)}_2^2 \le C_1\norm{\e}_2^2 + C_2\left( \norm{\alphabf - [\alphabf]_k}_2^2 + \frac{1}{{k}}\norm{\alphabf - [\alphabf]_k}_1^2 \right),
\end{eqnarray}
where $[\alphabf]_k$ is the best $k$-term approximation of $\alphabf$, and $C_1$ and $C_2$ are two constants depending on $\matr{A}$ and the algorithms\footnote{Note that \eqref{eq:synthesis_advers_noise_bound} is a generalization of the bound in \eqref{eq:stab_rec} for the case that $\alphabf$ is a non-exact $k$-sparse vector.} (See \cite{Candes06Near, Needell09CoSaMP, Dai09Subspace, Blumensath09Iterative, Foucart11Hard, Zhang11Sparse}); or (ii) for the case that $\vect{e}$ is a zero-mean white Gaussian   noise with variance $\sigma^2$, it is guaranteed that \rg{with a high probability,}
\begin{eqnarray}
\label{eq:synthesis_Gauss_noise_bound}
\norm{\alphabf - \RP(\y)}_2^2 \le C_3 k \sigma^2 \log(n) + C_4\left( \norm{\alphabf - [\alphabf]_k}_2^2 + \frac{1}{k}\norm{\alphabf - [\alphabf]_k}_1^2 \right),
\end{eqnarray}
where $C_3$ and $C_4$ are two constants depending on $\matr{A}$ and the algorithms (See  \cite{Candes07Dantzig, Bickel09Simultaneous, BenHaim09Coherence, Giryes12RIP}).

Assuming that $\OM$ in Problem $\cal{P}$ is a frame, we propose the following sampling and reconstruction strategy:
\begin{itemize}
\item Set the sensing matrix to be $\M = \matr{A} \OM$. In this case we have $\y = \M\x + \e = \matr{A}\OM\x + \e$ and therefore we can apply algorithm $\RP$ to recover $\OM\x$ as it is a $k$-sparse  (or approximately so) vector. 
\item Compute an estimate for $\OM\x$: $\hat{\alphabf} = \RP(\vect{y})$.
\item Use the frame's Moore-Penrose pseudo-inverse to recover $\x$: $\hat{\x} = \OM^\dag \hat\alphabf$.
\end{itemize}
This algorithm is summarized also in Algorithm~\ref{alg:frame_sampling}.
Remark that we sample in the transform domain of $\OM$, as we sample with $\M = \matr{A}\OM$, and then recover only with $\matr{A}$ the transform coefficients of $\x$, i.e. $\OM\x$. 
 Note also that in the final step, where we calculate $\hat{\x} = \OM^{\dag}\hat\w$, we may replace $\OM^\dag$ with any dictionary that satisfies $\D\OM = \I$.

\begin{algorithm}
\caption{Signal Recovery from Samples of Frames in the Transform Domain} \label{alg:frame_sampling}
\begin{algorithmic}

\REQUIRE $k$, \rg{$\matr{A} \in \RR{m \times n}$, $\OM \in \RR{n \times d}$},  $\y$, $\RP$, where $\vect{y} = \matr{A}\OM\vect{x}
+ \vect{e}$, $\OM\x$ is a $k$-sparse vector or approximately so, $\e$ is
an additive noise, and $\RP(\cdot) = \RP(\cdot|\matr{A},k)$ is a synthesis recovery program for $k$-sparse signals under the matrix $\matr{A}$.

\ENSURE $\hat{\vect{x}}$: Approximation of
$\vect{x}$.

\STATE Get a transform domain proxy for $\OM\x$: $\hat\w = \RP(\y|\matr{A},k)$

\STATE  Signal recovery: $\hat{\x} = \OM^{\dag}\hat\w$, generating a signal estimate using the transform domain proxy.

\end{algorithmic}
\end{algorithm}

The following theorem provide\rg{s} guarantees for signal recovery using the above scheme given that the synthesis reconstruction program used in it $\RP$ satisfies either \eqref{eq:synthesis_advers_noise_bound} or \eqref{eq:synthesis_Gauss_noise_bound}, or both of them.

\begin{thm}[Signal recovery from samples of frames in the transform domain]
\label{thm:sig_sample_frames}
Consider the problem $\cal P$ such that $\matr{M} = \matr{A}\OM$ and $\OM$ is a frame with a lower frame bound $A$. 
Let $\hat\x$ be the output of Algorithm~\ref{alg:frame_sampling} with the synthesis program $\RP(\cdot | \matr{A}, k)$.
If $\e$ is a bounded additive adversarial noise and \eqref{eq:synthesis_advers_noise_bound} holds for $\RP(\cdot | \matr{A}, k)$ then
\begin{eqnarray}
\label{eq:transform_advers_noise_bound}
&& \hspace{-0.3in} \norm{\x - \hat\x}_2^2
\le \frac{C_1}{A^2}\norm{\vect{e}}_2^2+
  \frac{C_2}{A^2}\left(\norm{\OM_{T^C}\x}_2^2 + \frac{1}{{k}}\norm{\OM_{T^C}\x}_1^2 \
   \right),
\end{eqnarray}
implying a stable recovery. 
If $\e$ is a zero-mean white Gaussian noise with variance $\sigma^2$  and \eqref{eq:synthesis_Gauss_noise_bound} holds for $\RP(\cdot | \matr{A},k)$ then \rg{with a high probability,}\footnote{\rg{Remark that it is also possible to provide guarantees for the expectation of the error, given a variant of \eqref{eq:synthesis_Gauss_noise_bound} that bounds the expectation of the error like in \cite{Candes07Dantzig, Giryes12RIP}}}.
\begin{eqnarray}
\label{eq:transform_Gauss_noise_bound}
&& \hspace{-0.3in} \norm{\x - \hat\x}_2^2
\le \frac{C_3}{A^2}k\log{n}\sigma^2+
  \frac{C_4}{A^2}\left(\norm{\OM_{T^C}\x}_2^2 + \frac{1}{{k}}\norm{\OM_{T^C}\x}_1^2 \
   \right),
\end{eqnarray}
implying a denoising effect. 
The constants $C_1, C_2, C_3, C_4$ are the same as in \eqref{eq:synthesis_advers_noise_bound} and \eqref{eq:synthesis_Gauss_noise_bound}.
\end{thm}
{\em Proof:}
We prove only the bound in \eqref{eq:transform_advers_noise_bound}. The proof for \eqref{eq:transform_Gauss_noise_bound} is very similar and omitted. Assume that \eqref{eq:synthesis_advers_noise_bound} holds. Then since $\y = \matr{A}\OM\x + \e$, we have that
\begin{eqnarray}
\label{eq:}
\norm{\OM\x - \hat{\w}}_2^2  = \norm{\OM\x - \RP(\y)}_2^2 \le C_1\norm{\e}_2^2 + C_2\left( \norm{\OM\x - [\OM\x]_k}_2^2 + \frac{1}{{k}}\norm{\OM\x - [\OM\x]_k}_1^2 \right).
\end{eqnarray}
We get \eqref{eq:transform_advers_noise_bound}  by using the facts that (i) $[\OM\x]_k = \OM_{T}\x$ and therefore $\OM\x - [\OM\x]_k = \OM_{T^C}\x$; (ii) $\OM$ is a frame with a lower frame bound $A$ and therefore $\norm{\OM^\dag}_2 \le \frac{1}{A}$; and (iii) $\x = \OM^\dag \OM\x$ and thus $\norm{ \x - \hat{\x} }_2 = \norm{\OM^\dag \left( \OM\x - \hat{\w} \right)}_2$.
\hfill $\Box$
\bigskip

\rg{This theorem provides the same guarantees derived for analysis algorithms, which were designed especially for the analysis framework, using already existing methods from the synthesis model. The wide use of the latter and the large variety of programs available for it allow  recovering a signal from a small number of measurements with more ease, using our new sampling scheme. In addition, we may say that the above theorem demonstrates that our new sampling scheme allows transferring almost any existing result from the synthesis framework to the analysis one. 
One example is the ability to set $\matr{A}$ to be an expander graph. In this case, it is possible to recover the signal $\x$ using only $k$ steps \cite{Jafarpour09Efficient}. To the best of our knowledge, such an efficient strategy does not exist for the analysis framework. }

\subsection{Guarantees for the 2D-DIF Operator}
\label{sec:TV_guarantees}

Having a guarantee for frames we turn to provide a guarantee for 2D-DIF, the two-dimensional finite difference operator. For convenience we assume that $\x$ is an image (column stacked) of size $N \times N = d$ ($N= \sqrt{d}$). 
Notice that unlike frames, for the 2D-DIF operator a small distance in the transform domain does not imply a small distance in the signal domain. For example, the distance between two constant images is zero in the transform domain of the 2D-DIF operator. However, it can be arbitrarily as large as we want depending on the constant value we assign to each image. 
Therefore, it is impossible to recover a signal by just using the scheme we have in Algorithm~\ref{alg:frame_sampling}.
Note that the problem lies in the last stage of the algorithm as we do not have enough information to get back stably from the transform domain to the signal domain.  
Note that also if we will add rows to $\OM_{2D-DIF}$ and then apply a pseudo inverse, we will not have a stable recovery in the signal domain given the recovery in the transform domain (See \cite{Needell13Stable, Needell13Near} for more details).

Therefore we utilize the tools used in \cite{Needell13Stable} that studies  the performance of the 2D-DIF operator with the analysis $\ell_1$-minimization, which is known also as the anisotropic total variation (TV).
Two key steps are used in that work for developing the result for TV:
\begin{itemize}
\item The construction of the measurements: 
\begin{eqnarray}
\y = \left( \begin{array}{c}
\M_1 \x_{nfr} \\
\M_1 \x_{nlr} \\
\M_2 \x_{nfc} \\
\M_2 \x_{nlc} \\
\rg{\matr{B}} \x
\end{array}
\right) + \e,
\end{eqnarray}
where $\x_{nfr}$, $\x_{nlr}$, $\x_{nfc}$ and $\x_{nlc}$ are versions of $\x$ with no 
first row, last row, first column or last column respectively. In addition, $\M_1, \rg{\M_2}  \in \RR{m_1 \times N(N-1)}$ are assumed to satisfy the RIP with $\delta_{5k} < \frac{1}{3}$ and $\rg{\matr{B}}\matr{H}^{-1}$ is assumed to satisfy the RIP with $\delta_{2k} < 1$, where $\matr{H}$ is the bivariate Haar transform and \rg{$\matr{B} \in \RR{m_2 \times d}$}.
\item The usage of the relationship between $\OM_{2D-DIF}$ and $\matr{H}$: For any vector $\v$, if $\norm{\OM_{2D-DIF}\v}_0 \le k$ then $\norm{\matr{H}\v}_0 \le k \log(d)$.
\end{itemize}

The first two measurement matrices $\M_1$ and $\M_2$ provide information about the derivatives of $\x$ and lead to a stable recovery of $\OM_{2D-DIF}\x$, the discrete gradient vector of $\x$. 
As we have mentioned before $\OM_{2D-DIF}$ is non-invertible. Therefore, the reconstruction of the derivatives is not enough for recovering the signal. For this purpose the third matrix $\rg{\matr{B}}$ is used to guarantee stable recovery also in the signal domain. This is achieved using the following theorem:

\begin{thm}[Strong Sobolev inequality. Theorem~8 in \cite{Needell13Stable}]
\label{thm:strong_sobolev}
Let $N$ be a power of $2$ and $\rg{\matr{B}}$ be a linear map which, composed with the inverse bivariate Haar transform \rg{$\matr{B}\matr{H}^{-1} \in \RR{m_2 \times d}$}, has the RIP with a constant $\delta_{2k}<1$. Suppose that for $\vect{z} \in \RR{d}$ we have $\norm{\rg{\matr{B}}\vect{z}}_2 \le \epsilon$. Then
\begin{eqnarray}
\norm{\vect{z}}_2 \le \frac{2C_{\matr{H}}}{1-\delta_{2k}}\frac{1}{\sqrt{k}}\log(d/k)\norm{\OM_{2D-DIF}\vect{z}}_1 + \frac{1}{1-\delta_{2k}}\epsilon,
\end{eqnarray} 
where $C_{\matr{H}} =36(480\sqrt{5} + 168\sqrt{3})$. 
\end{thm}

We utilize the above theorem for extending our sampling technique for the 2D-DIF operator. 
By observing again the samples generated by $\M_1$ and $\M_2$, and denoting by $\OM_v$ and $\OM_h$ the vertical and horizontal difference of $\OM_{2D-DIF}$ respectively, we can write 
$\M_1\x_{nfr} - \M_1\x_{nlr} = \M_1\OM_{v}\x$ and $\M_1\x_{nfc} - \M_1\x_{nlc} = \M_2\OM_{h}\x$. Alternatively, we can rewrite it as
\begin{eqnarray}
\left(\begin{array}{cc}
\M_1 & 0 \\
0 & \M_2
\end{array} \right) \OM_{2D-DIF}\x,
\end{eqnarray} 
and we end up with having samples from the derivatives domain. 
Notice that we do not have to restrict ourselves to a block diagonal matrix composed of two linear maps for sampling each derivative direction. We can use any sampling operator that has recovery guarantees in the synthesis framework for reconstructing the coefficients in the transform domain. We denote this reconstruction by $\hat\w$.

In order to recover the signal from its proxy $\hat\w$, we take more measurements of the original signal $\x$. These are taken using a matrix $\matr{B}$ for which $\matr{B}\matr{H}^{-1}$ (its composition with the inverse bivariate Haar transform) has the RIP with a constant $\delta_{2k} < 1$. Given these measurements, $\y_2 = \matr{B} \x + \e_2$, we get a recovery of the signal by solving 
 \begin{eqnarray}
\label{eq:2D_DIF_signal_recovery}
\hat{\x}_{\text{2D-DIF}} = \argmin_{\tilde{\x}}\norm{\OM\tilde{\x} - \hat{\w}_{\text{2D-DIF}}}_1
& s.t. & \norm{\matr{B}\tilde{\x} - \y_2}_2  \le \norm{\e_2}_2.
\end{eqnarray}

\begin{algorithm}
\caption{Signal Recovery from Samples of 2D-DIF in the Transform Domain} \label{alg:2D_DIF_sampling}
\begin{algorithmic}

\REQUIRE $k$, $\rg{\matr{A} \in \RR{m_1 \times n}}$, $\matr{B} \in \RR{m_2 \times d}$,  $\OM_{2D-DIF} $, $\y, \RP$, where $\y =\left[ \begin{array}{c}
\y_1 \\ \y_2
\end{array} \right]$ such that $\y_1 = \matr{A}\OM_{2D-DIF}\vect{x}
+ \vect{e}_1$ and $\y_2  = \matr{B}\x +e_2$, $\OM_{2D-DIF}\x$ is $k$ sparse or approximately so, $\e = \left[ \begin{array}{c}
\e_1 \\ \e_2
\end{array} \right]$ is
an additive noise, and $\RP(\cdot) = \RP(\cdot,\matr{A},k)$ is a synthesis recovery program for $k$-sparse representation under the matrix $\matr{A}$.

\ENSURE $\hat{\vect{x}}$: Approximation of
$\vect{x}$.

\STATE Get a transform domain proxy for $\OM\x$: $\hat\w = \RP(\y_1,\matr{A},k)$

\STATE  Signal recovery: Calculate $\hat\x$ using \rg{\eqref{eq:2D_DIF_signal_recovery}} with $\vect{y}_2$ and $\hat{\w}$.

\end{algorithmic}
\end{algorithm}

To sum it up, our sampling strategy for the 2D-DIF operator consists of taking two sets of measurements. The first in the transform domain, $\y_1= \matr{A}\OM_{2D-DIF}\x + \e_1$, leads to reconstruction of the gradient components. The second is taken with a linear map which is well behaved if applied together with the inverse of the bivariate Haar, $\y_2 = \matr{B}\x + \e_2$, where its sole purpose is to convert the transform domain estimate into a signal estimate using \eqref{eq:2D_DIF_signal_recovery}. 
Note that the linear map we use for sampling is  $\M =\left[ \begin{array}{c}
\matr{A}\OM_{2D-DIF} \\  \matr{B}
\end{array} \right]$ and our measurements are of the form $\y = \M \x + \e$, where $\e = \left[ \begin{array}{c}
\e_1 \\ \e_2
\end{array} \right]$. 
Our recovery strategy from these samples is summarized in Algorithm~\ref{alg:2D_DIF_sampling}. Note that in \eqref{eq:2D_DIF_signal_recovery} we can use $\norm{\e}_2$ instead of $\norm{\e_2}_2$ if we do not have a good bound for the latter.

For the theoretical study of Algorithm~\ref{alg:2D_DIF_sampling} we make a different assumption on the used synthesis program $\RP$. Instead of the bounds in \eqref{eq:synthesis_advers_noise_bound} and \eqref{eq:synthesis_Gauss_noise_bound} we assume that the following holds: 
\begin{eqnarray}
\label{eq:synthesis_advers_noise_bound_l1}
\norm{\alphabf - \RP(\y)}_1 \le C_5\sqrt{k}\norm{\e}_2 + C_6 \norm{\alphabf - [\alphabf]_k}_1.
\end{eqnarray}
Such a bound holds for the synthesis $\ell_1$-minimization with RIP matrices \cite{Needell13Stable}.
With this assumption we are ready to introduce the recovery guarantee for Algorithm~\ref{alg:2D_DIF_sampling}.

\begin{thm}[Stable signal recovery from samples of 2D-DIF in the transform domain]
\label{thm:TV_adversarial_noise}
Consider the problem $\cal P$ such that $\M =\left[ \begin{array}{c}
\matr{A}\OM_{2D-DIF} \\  \matr{B}
\end{array} \right]$, \rg{where $\matr{A}$ has the RIP with a constant $\delta_{ak}$ for a certain constant $a \ge 1$,} $\OM_{2D-DIF}$ is the 2D-DIF operator and $\matr{B}\matr{H}^{-1}$ has the RIP with a constant $\delta_{2k} <1$.
Let $\hat\x$ be the output of \rg{Algorithm~\ref{alg:2D_DIF_sampling}} with the synthesis program $\RP(\cdot | \matr{A}, k)$.
If $\e$ is a bounded additive adversarial noise and \eqref{eq:synthesis_advers_noise_bound_l1} holds for $\RP(\cdot | \matr{A}, k)$ then 
\begin{eqnarray}
\norm{ \hat\x - \x}_2 \le \log(d/k)\left(C_7\norm{\e}_2 + \frac{C_8 }{\sqrt{k}}\norm{\OM_{T^C}\x}_1 \right),
\end{eqnarray}
implying a stable recovery, where $C_7$ and $C_8$ are functions of $C_{\matr{H}}$ and $\delta_{2k}$.
\end{thm}
{\em Proof:}
Since $\hat\x$ is a minimizer of \rg{\eqref{eq:2D_DIF_signal_recovery}} we have that
\begin{eqnarray}
\norm{\matr{B}\hat\x - \y_2}_2  \le \norm{\vect{e}_2}_2.
\end{eqnarray} 
Since $\y_2 = \matr{B}\x + \e_2$ we have from the triangle inequality that 
\begin{eqnarray}
\norm{\matr{B}(\hat\x - \x)}_2 \le 2\norm{\e_2}_2.
\end{eqnarray}
Therefore, setting $\vect{z} = \hat\x - \x$ in Theorem~\ref{thm:strong_sobolev} we have
\begin{eqnarray}
\label{eq:hatx_x_2_first}
\norm{ \hat\x - \x}_2 \le \frac{2C_{\matr{H}}}{1-\delta_{2k}}\frac{1}{\sqrt{k}}\log(d/k)\norm{\OM_{2D-DIF}( \hat\x - \x)}_1 + \frac{2}{1-\delta_{2k}}\norm{\e_2}_2.
\end{eqnarray} 
From the triangle inequality we have 
\begin{eqnarray}
\label{eq:OM_2DDIF_hatx_x}
\norm{\OM_{2D-DIF}( \hat\x - \x)}_1 \le \norm{\OM_{2D-DIF} \hat\x - \hat\w}_1 + \norm{\hat\w - \OM_{2D-DIF}\x}_1.
\end{eqnarray}
Since $\x$ is a feasible solution to \rg{\eqref{eq:2D_DIF_signal_recovery}} and $\hat\x$ is its minimizer we have
\begin{eqnarray}
\label{eq:OM_2DDIF_hatx_w}
 \norm{\OM_{2D-DIF}\hat\x - \hat\w}_1 \le  \norm{\OM_{2D-DIF} \x - \hat\w}_1.
\end{eqnarray}
Plugging \eqref{eq:OM_2DDIF_hatx_w} in \eqref{eq:OM_2DDIF_hatx_x} we have 
\begin{eqnarray}
\label{eq:OM_2DDIF_hatx_x_2}
\norm{\OM_{2D-DIF}( \hat\x - \x)}_1 \le  2\norm{\OM_{2D-DIF}\x - \hat\w}_1. 
\end{eqnarray}
Notice that we can bound the right hand side (rhs) of \eqref{eq:OM_2DDIF_hatx_x_2} with \eqref{eq:synthesis_advers_noise_bound_l1}, where $\alphabf = \OM\x$ and $\hat\alphabf = \hat\w$. 
Therefore, by combining \eqref{eq:OM_2DDIF_hatx_x_2} and \eqref{eq:synthesis_advers_noise_bound_l1} with \eqref{eq:hatx_x_2_first} we have
\begin{eqnarray}
\norm{ \hat\x - \x}_2 \le \frac{2C_{\matr{H}}}{1-\delta_{2k}}\log(d/k)\left(C_5\norm{\e}_2 + \frac{C_6 }{\sqrt{k}}\norm{\OM_{T^C}\x}_1 \right) + \frac{2}{1-\delta_{2k}}\norm{\e_2}_2.
\end{eqnarray}

\hfill $\Box$
\bigskip

\begin{rem}
\rg{An example of a procedure $\RP(\cdot | \matr{A}, k)$, for which  \eqref{eq:synthesis_advers_noise_bound_l1} holds, is the synthesis  $\ell_1$-minimization with $\matr{A}$ having the RIP with a constant $\delta_{5k} \le \frac{1}{3}$ \cite{Needell13Stable}}.
\end{rem}

\subsection{Guarantees for a General Analysis Operator}
\label{sec:general_guarantees}

\begin{algorithm}
\caption{Signal Recovery from Samples of a General Analysis Operator in the Transform Domain} \label{alg:general_sampling}
\begin{algorithmic}

\REQUIRE \rg{$k$, $\matr{A} \in \RR{m_1 \times n}$, $\matr{B} \in \RR{m_2 \times d}$,  $\OM$, $\y, p, \RP$, where $\y =\left[ \begin{array}{c}
\y_1 \\ \y_2
\end{array} \right]$ such that $\y_1 = \matr{A}\OM\vect{x}
+ \vect{e}_1$ and $\y_2  = \matr{B}\x +\e_2$, $\OM\x$ is $k$ sparse or approximately so, $\e = \left[ \begin{array}{c}
\e_1 \\ \e_2
\end{array} \right]$ is
an additive noise, $p$ is the $\ell_p$ norm used in this algorithm, and $\RP(\cdot) = \RP(\cdot,\matr{A},k)$ is a synthesis recovery program for $k$-sparse representation under the matrix $\matr{A}$.}

\rg{\ENSURE $\hat{\vect{x}}$: Approximation of
$\vect{x}$.}

\rg{\STATE Get a transform domain proxy for $\OM\x$: $\hat\w = \RP(\y_1,\matr{A},k)$.}

\rg{\STATE  Signal recovery:
 \begin{eqnarray}
\label{eq:general_signal_recovery}
\hat{\x} = \argmin_{\tilde{\x}}\norm{\OM\tilde{\x} - \hat{\w}}_p
& s.t. & \norm{\matr{B}\tilde{\x} - \y_2}_2  \le \norm{\e_2}_2.
\end{eqnarray} }

\end{algorithmic}
\end{algorithm}

Extending this idea further we do not restrict the sampling strategy in Algorithm~\ref{alg:2D_DIF_sampling} only to  
$\OM_{2D-DIF}$. \rg{We present this extension in Algorithm~\ref{alg:general_sampling}.}
It can be applied for any operator for which a stable recovery in the coefficients domain implies a stable recovery in the signal domain by some additional measurements of the signal. 
\rg{The following theorem, which is similar to Theorem~\ref{thm:TV_adversarial_noise},
 provides a recovery guarantee for this generalized scheme.}

\begin{thm}[Stable signal recovery from samples of a general analysis operator  in the transform domain]
\label{thm:general_adversarial_noise}

\rg{Consider the problem $\cal P$ such that $\M =\left[ \begin{array}{c}
\matr{A}\OM \\  \matr{B}
\end{array} \right]$, where $\matr{A} \in \RR{m_1 \times n}$ and $\OM$ is a general analysis operator.
Suppose $\matr{B}$ is a matrix such that for any $\z \in \RR{d}$,
$\norm{{\matr{B}}\vect{z}}_2 \le \epsilon$ implies
\begin{eqnarray}
\label{eq:z_general_beta_gamma}
\norm{\vect{z}}_2 \le \beta\norm{\OM\vect{z}}_p + \gamma\epsilon.
\end{eqnarray}
and that for any $\alphabf \in \RR{n}$ and $\y_1 \in \RR{m_1}$
\begin{eqnarray}
\label{eq:synthesis_advers_noise_bound_general}
\norm{\alphabf - \RP(\y)}_p \le \zeta\norm{\e}_2 + \xi \norm{\alphabf - [\alphabf]_k}_1.
\end{eqnarray}
holds for the synthesis program $\RP(\cdot | \matr{A}, k)$.
Let $\hat\x$ be the output of Algorithm~\ref{alg:general_sampling} with the program $\RP(\cdot | \matr{A}, k)$ and $\e$ be a bounded additive adversarial noise. Then
\begin{eqnarray}
\label{eq:general_adversarial_noise_bound}
\norm{ \hat\x - \x}_2 \le 2\beta\left(\zeta\norm{\e}_2 + \xi\norm{\OM_{T^C}\x}_1 \right) + 2\gamma\norm{\e_2}_2.
\end{eqnarray} }
\end{thm}
{\em Proof:}
\rg{As the proof is very similar to the one of Theorem~\ref{thm:TV_adversarial_noise} we present it briefly.
Using the same steps that led to \eqref{eq:hatx_x_2_first} and \eqref{eq:OM_2DDIF_hatx_x_2} we have 
\begin{eqnarray}
\label{eq:hatx_x_2_first_general}
\norm{ \hat\x - \x}_2 \le 
2\beta\norm{\OM(\hat\x - \x)}_p + 2\gamma\norm{\e_2}_2.
\end{eqnarray} 
and
\begin{eqnarray}
\label{eq:OM_hatx_x_2}
\norm{\OM( \hat\x - \x)}_p \le  2\norm{\OM\x - \hat\w}_p. 
\end{eqnarray}
Plugging \eqref{eq:synthesis_advers_noise_bound_general} in 
 \eqref{eq:OM_hatx_x_2}, with $\alphabf = \OM\x$, and then combining the result with \eqref{eq:hatx_x_2_first_general} lead to \eqref{eq:general_adversarial_noise_bound}.}
\hfill $\Box$
\bigskip

\rg{Notice that the result in Theorem~\ref{thm:TV_adversarial_noise} is a special case of the above theorem. We present two other special cases in the following two corollaries.
The first is a generalization of Theorem~\ref{thm:TV_adversarial_noise} for $L$-dimensional signals and the $L$D-DIF operator, the $L$ dimensional finite difference analysis dictionary.}

\begin{cor}[Stable signal recovery from samples of $L$D-DIF in the transform domain]
\label{cor:TVN_adversarial_noise}
\rg{Consider the problem $\cal P$ such that $\M =\left[ \begin{array}{c}
\matr{A}\OM_{LD-DIF} \\  \matr{B}
\end{array} \right]$, where $\OM_{LD-DIF}$ is the $L$D-DIF operator, $\matr{B}\matr{H}^{-1}$ has the RIP with a constant $\delta_{2k} <1$, and $\matr{H}$ is the $L$-dimensional Haar wavelet transform.
Let $\hat\x$ be the output of Algorithm~\ref{alg:general_sampling} with  the synthesis program $\RP(\cdot | \matr{A}, k)$ and $p=1$.
If $\e$ is a bounded additive adversarial noise and \eqref{eq:synthesis_advers_noise_bound_l1} holds for $\RP(\cdot | \matr{A}, k)$ then 
\begin{eqnarray}
\norm{ \hat\x - \x}_2 \le \log(d)\left(C_9\norm{\e}_2 + \frac{C_{10} }{\sqrt{k}}\norm{\OM_{T^C}\x}_1 \right),
\end{eqnarray}
implying a stable recovery, where $C_9$ and $C_{10}$ are certain constants.}
\end{cor}

\rg{The proof follows from a generalized version of Theorem~\ref{thm:strong_sobolev} for the $L$-dimensional case (Theorem~6 in \cite{Needell13Near}) that provides \eqref{eq:z_general_beta_gamma} with $\gamma =1 $ and $\beta = \log(d)\frac{C}{\sqrt{k}}$, where $C$ is a certain constant.
}

\begin{rem}
\rg{Notice that one may further generalize Theorem~\ref{thm:general_adversarial_noise} to deal also with block sparsity \cite{stojnic2009reconstruction, eldar2009robust, tropp2006algorithms, baraniuk2010model},
i.e., the case that $\OM_1\vect{x}, \OM_2\vect{x}, \dots, \OM_L\vect{x}$ are jointly sparse, where $\OM = \left[\OM_1^T, \OM_2^T, \dots, \OM_L \right]^T$. In this case, the $\ell_1$-norm applied on vectors in $\RR{n}$ in Algorithm~\ref{alg:general_sampling}, Theorem~\ref{thm:general_adversarial_noise} and \eqref{eq:synthesis_advers_noise_bound_l1}
needs to be replaced with the mixed $\ell_{1,2}$-norm\footnote{Applying an $\ell_2$-norm on the rows followed by an $\ell_1$-norm on the resulted vector.} applied on matrices in $\RR{\frac{n}{L} \times L}$. 
An example for such a case is the $L$-dimensional isotropic total variation, where $\OM_i$ is the derivative in the $i$-th dimension (if $L = 2$ then $\OM_1$ and $\OM_2$ are the  horizontal and vertical derivatives respectively). 
Note that it can be shown that the $\ell_{1,2}$-minimization algorithm satisfies  a version of \eqref{eq:synthesis_advers_noise_bound_l1} with the $\ell_{1,2}$-norm. In addition, Theorem~6 in \cite{Needell13Near} provides a bound in the form of \eqref{eq:z_general_beta_gamma} with the $\ell_{1,2}$-norm instead of the $\ell_1$-norm. Therefore, it is possible to derive a theorem similar to Corollary~\ref{cor:TVN_adversarial_noise} equivalent to the theorems for the isotropic TV in \cite{Needell13Near}.
We leave the details to the interested reader. }
\end{rem}

\rg{The second corollary considers operators that can be viewed as part of a frame.}

\begin{cor}[Stable signal recovery from samples of a partial frame]
\label{cor:partial_frame}
\rg{Consider the problem $\cal P$ such that $\M =\left[ \begin{array}{c}
\matr{A}\OM \\  \matr{B}
\end{array} \right]$, where $\OM$ is a matrix for which there exists $\tilde{\OM}$ such that $\OM_F = \left[\OM^T, \tilde{\OM}^T \right]^T$ is a frame with a lower frame bound $A$, and $\sigma_{min}\left(\matr{B}\tilde{\OM}^\dag \right) \ge C_{11}$ and  $\norm{\matr{B}\left(\matr{I} - \tilde{\OM}^{\dag}\tilde{\OM}\right)} \le C_{12}$ for constants $C_{11}$ and $C_{12}$ satisfying $\frac{C_{12}}{C_{11} A} < 1$.
Let $\hat\x$ be the output of Algorithm~\ref{alg:general_sampling} with  the synthesis program $\RP(\cdot | \matr{A}, k)$ and $p = 2$.
If $\e$ is a bounded additive adversarial noise and \eqref{eq:synthesis_advers_noise_bound_general} holds for $\RP(\cdot | \matr{A}, k)$ with $\zeta = C_{13}$ and $\xi = \frac{C_{14}}{\sqrt{k}}$, where $C_{13}$ and $C_{14}$ are certain constants, then 
\begin{eqnarray}
\norm{ \hat\x - \x}_2 \le \left(C_{15}\norm{\e}_2 + \frac{C_{16} }{\sqrt{k}}\norm{\OM_{T^C}\x}_1 \right),
\end{eqnarray}
implying a stable recovery, where $C_{15}$ and $C_{16}$ are constants dependent only on $A$, $C_{11}$, $C_{12}$, $C_{13}$ and $C_{14}$.}
\end{cor}

{\em Proof:}
\rg{ For the proof we just need to show that \eqref{eq:z_general_beta_gamma} holds. Using the lower frame bound followed by the triangle inequality and the fact that $\sigma_{min}\left(\matr{B}\tilde{\OM}^\dag \right) \ge C_{11}$, we have
\begin{eqnarray}
\label{eq:partial_frame_proof_step1}
\norm{\z}_2 \le \frac{1}{A}\norm{\OM_{F}\z}_2 \le \frac{1}{A}\norm{\OM\z}_2 +\frac{1}{A}\norm{\tilde{\OM}\z }_2  \le \frac{1}{A}\norm{\OM\z}_2 +\frac{1}{AC_{11}}\norm{\matr{B}\tilde{\OM}^\dag\tilde{\OM}\z }_2
\end{eqnarray}
Using the triangle inequality and the fact that $\norm{\matr{B}\left(\matr{I} - \tilde{\OM}^{\dag}\tilde{\OM}\right)} \le C_{12}$ we have
\begin{eqnarray}
\label{eq:partial_frame_proof_step2}
\norm{\matr{B}\tilde{\OM}^\dag\tilde{\OM}\z }_2 \le
\norm{\matr{B}\z }_2 + C_{12}\norm{\z }_2 
\end{eqnarray}
Plugging \eqref{eq:partial_frame_proof_step2} in \eqref{eq:partial_frame_proof_step1} with some simple arithmetical steps lead to
\begin{eqnarray}
\left(1 - \frac{C_{12}}{AC_{11}} \right)\norm{\z}_2 \le  \frac{1}{A}\norm{\OM\z}_2 +\frac{1}{AC_{11}}\norm{\matr{B}\z }_2.
\end{eqnarray}  
Notice that by the assumptions of the corollary $1 - \frac{C_{12}}{AC_{11}} >0$. This equation provides the constants in \eqref{eq:z_general_beta_gamma}, completing the proof.}
\hfill $\Box$
\bigskip

\begin{rem}
\rg{An example for a program $\RP(\cdot | \matr{A}, k)$ that satisfies the assumption of the theorem is CoSaMP \cite{Needell09CoSaMP}.}
\end{rem}
\begin{rem}
\rg{An example for a matrix $\matr{B}$ that satisfies the assumptions of the theorem is $\matr{B} = \tilde{\OM}$. In this case $C_{11}= 1$ and $C_{12} = 0$.}
\end{rem}

\rg{Another family of analysis operators that might be of interest is the one of convolutional operators \cite{Hawe13Analysis}. In this case the condition number of $\OM$ is usually very large and the sampling strategy used with Algorithm~\ref{alg:general_sampling} is needed, as we cannot sample directly from the transform domain like in the case of frames. We leave the exploration of this case to a future research.}


\section{Epilogue - Do We Still Need Analysis Algorithms? }
\label{sec:exp}

Following the fact that our proposed recovery guarantees are similar to the ones achieved for the existing analysis algorithms and that sampling in the manifold dimension of analysis signals lead to unstable recovery \cite{Giryes14effective}, one may ask whether there is a need at all for reconstruction strategies that rely on the analysis model. For this reason we perform several experiments to compare the empirical recovery performance of our new sampling scheme, with synthesis $\ell_1$-minimization, and the standard sampling scheme, with analysis $ell_1$-minimization, for signals from the analysis framework. \rg{The minimizations are performed using {\em cvx} \cite{cvx, gb08}.} 

We start with the case of signals that are sparse after applying randomly generated tight-frames. We set $\OM \in \RR{144 \times 120}$, where the signal dimension is $d= 200$, and $k = 144 - 110$ (setting the signal intrinsic dimension to be $10$, see \cite{Giryes14GreedySignal} for more details).
In the standard sampling setup, the entries of the sensing matrix $\matr{M} \in \RR{d\gamma \times d}$, where $\gamma \in \{0.05, 0.1, 0.15, \dots, 1  \}$, are randomly generated from an i.i.d random Gaussian distribution, followed by a normalization of each column to have a unit $\ell_2$-norm. For the new scheme we set $\matr{M} = \matr{A}\OM$ with $\matr{A} \in \RR{d\gamma \times 1.2d}$ a random Gaussian matrix selected in the same way that $\matr{M}$ is selected in the standard sampling scheme.
For each value of $\gamma$ we generate $1000$ different sensing matrices and signals $\x$ that have sparsity $k$ under $\OM$. The signals are generated by projecting a randomly selected Gaussian vector to the subspace orthogonal to randomly selected $ n - k$ rows from $\OM$, followed by normalization of the vector.

\begin{figure*}[!t]
{\subfigure[Noiseless Case Recovery Rate]{\includegraphics[width=3in]{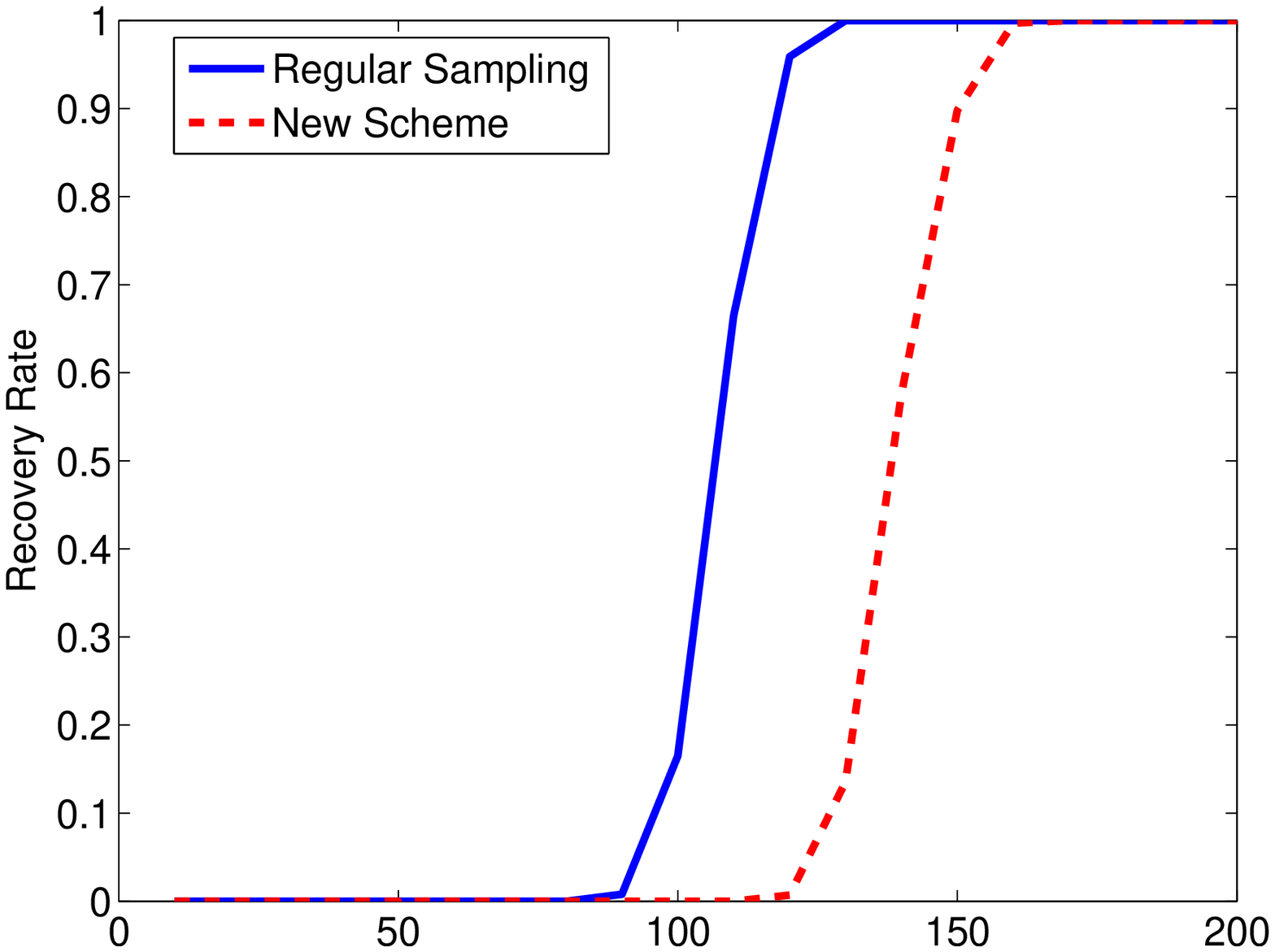}\label{fig:l1_rec_rate}}
\hfil
\subfigure[Noisy Case Mean Squared Error]{\includegraphics[width=3in]{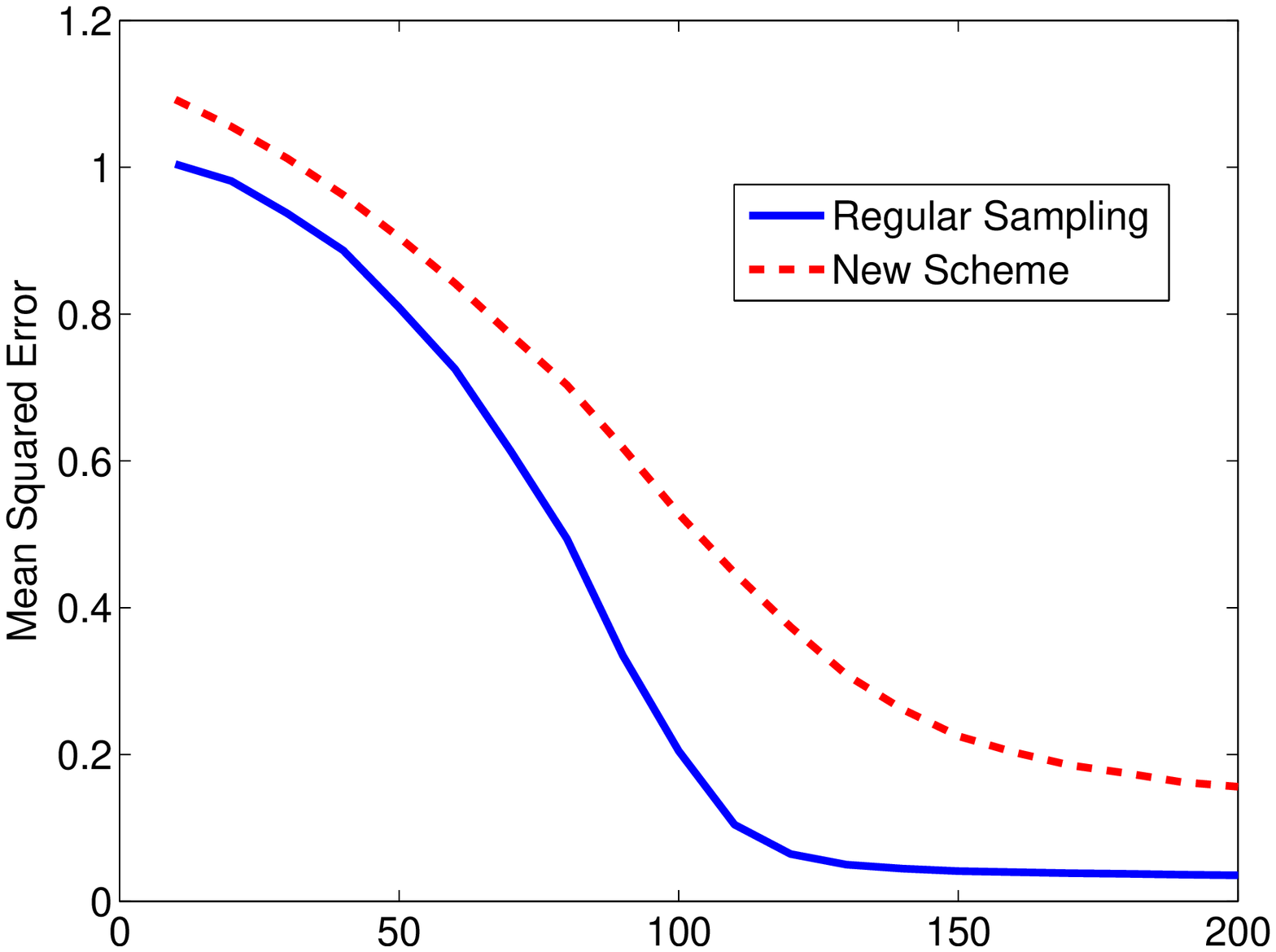}\label{fig:l1_error_rate}}%
}
\caption{Comparison between recovering a signal that belongs to the analysis model, with a frame as the analysis operator, using the standard sampling scheme with analysis algorithm and our new sampling scheme with synthesis algorithm. Left: Recovery rate for the noiseless case. Right: Reconstruction mean squared error in the noisy case.}
\label{fig:l1_frame}
\end{figure*}

\begin{figure*}[!t]
{\subfigure[Noiseless Case Recovery Rate]{\includegraphics[width=3in]{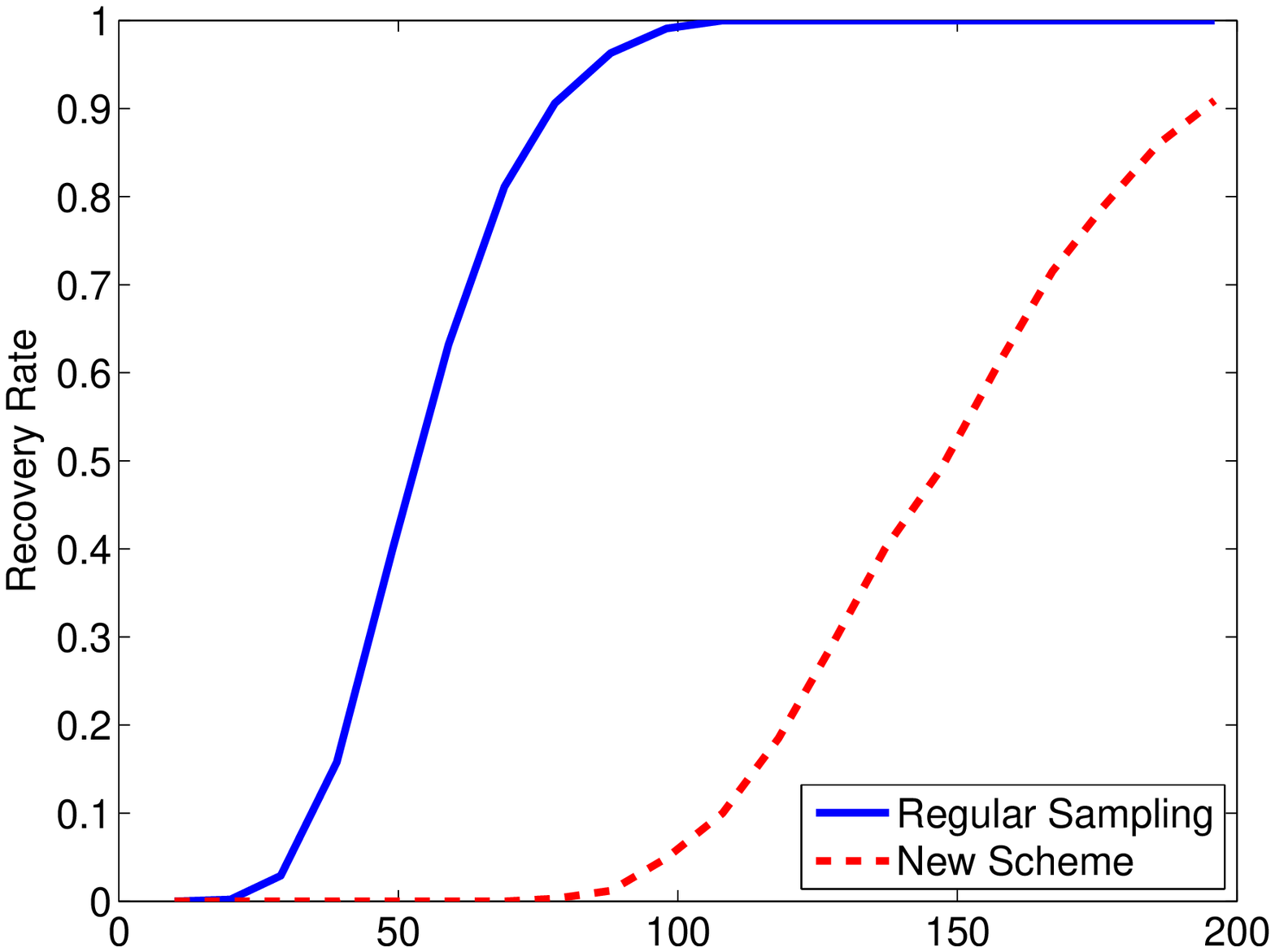}\label{fig:l1_error_rate_TV}}
\hfil
\subfigure[Noisy Case Mean Squared Error]{\includegraphics[width=3in]{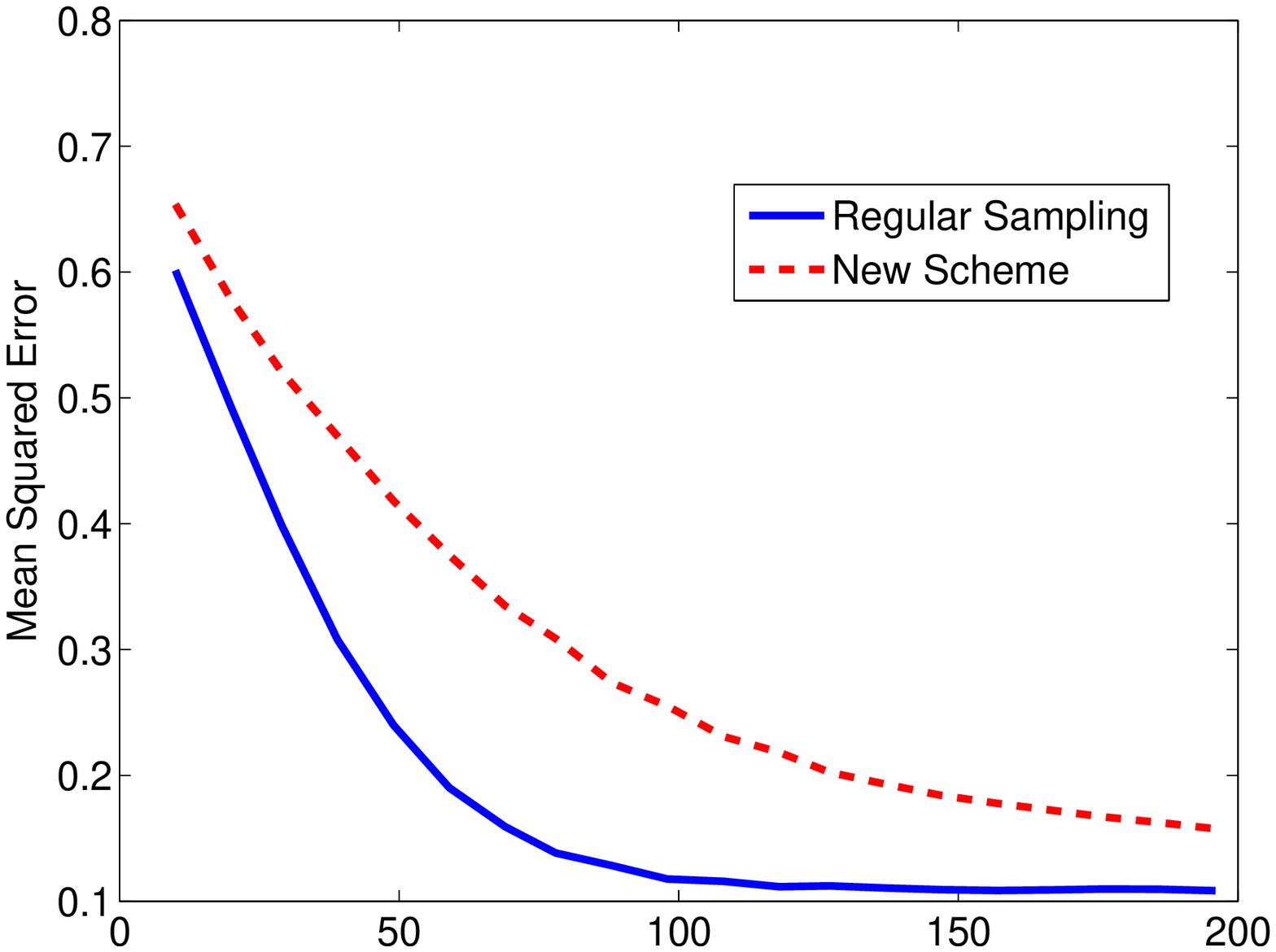}\label{fig:l1_error_td_TV}}%
}
\caption{Comparison between recovering a signal that belongs to the analysis model, with the 2D-DIF as the analysis operator, using the standard sampling scheme with analysis algorithm and our new sampling scheme with synthesis algorithm. Left: Recovery rate for the noiseless case. Right: Reconstruction mean squared error in the noisy case.}
\label{fig:l1_TV}
\end{figure*}

In Fig.~\ref{fig:l1_frame} we \rg{present} the recovery rate of the two algorithms in the noiseless and noisy cases. The noise is set to be i.i.d white Gaussian with $\sigma = 0.01$.
It can be seen that it is possible to recover signals from the analysis model using Algorithm~\ref{alg:frame_sampling}. However, this comes at the cost of using more samples in order to achieve the same recovery rate and error. 
This shows us that though the theoretical guarantees of the analysis algorithms take into account only $k$ and not the intrinsic dimension of the signals, losing the information about the latter, which happens when we sample in the transform domain, may harm the recovery. On the other hand, if we can  afford having more measurements, then we have the privilege of using existing synthesis algorithms, \rg{which have a large variety of efficient implementations compared to what is available for the analysis model.  For example, compare the methods available for the generic synthesis $\ell_1$-minimization problem \cite{Elad07wide, Hale08Fixed, Yin08Bregman, Wright09Sparse, Berg09Probing, Beck09Fast, Yang11Alternating,Becker11NESTA, Bach11Optimization, Treister12Multilevel} to the ones designed for the generic analysis $\ell_1$-minimization \cite{Zhao14Smoothing,Gu14Reconstruction}.
Remark that the advantage in efficiency is not unique to the $\ell_1$-relaxation alone.}
For more examples, \rg{we mention the sampling with expander graphs \cite{Jafarpour09Efficient} that does not have a counterpart in the analysis framework and} refer the reader to compare OMP with GAP \cite{Nam12Cosparse} or the synthesis greedy-like algorithms with their analysis versions \cite{Giryes14GreedySignal}.

We repeat the experiment with the 2D-DIF operator and compare analysis $\ell_1$-minimization with the scheme in Algorithm~\ref{alg:2D_DIF_sampling} that uses synthesis $\ell_1$-minimization for recovery. The signals we generate are random $14 \times 14$ images with four connected components. We start with a constant image and then add to it three additional connected components using a random walk on the image using the same technique in \cite{Giryes14effective}.
The sensing matrices are selected as in the previous experiment, where in the new sampling scheme we assign $2$ measurements (from the total number of measurements we use) in the noiseless case for the signal recovery from the transform domain proxy and $m/10$ in the noisy case.

Figure~\ref{fig:l1_TV} presents the reconstruction rate in the noiseless case and the recovery error in the noisy case, where the noise is the same as in the previous experiments. 
We see the same phenomenon that we saw in the previous experiment but stronger. As the redundancy the in analysis operator is bigger in this experiment, the number of measurements we need for the new scheme is relatively larger and the recovery error in the noisy case is higher. Another reason, other than the bigger redundancy, for the inferior performance in this case is that we separate the measurements we have into two parts, where in the standard scheme the analysis $\ell_1$-minimization uses all the measurements at once for the recovery of the signal. Note that this causes that even in the case that $m=d$ we do not get $100\%$ recovery. Clearly in this case we will just invert the measurement matrix instead of using neither of the two schemes.

\section{Discussion and Conclusion}
\label{sec:conc}

In this work we have presented a new sampling and recovery strategy for signals that are sparse under frames or the 2D-DIF operator in the analysis model. Our scheme utilizes existing algorithms from the synthesis sparsity model to recover signals that belong to the analysis framework. 
The advantage of this technique is that it enables the usage of existing tools for recovering signals from another model. Though in theory there is no additional cost for the usage of this scheme, it seems that in practice its advantage comes at the cost of the usage of more measurements in the sampling stage.
This gap between the theory and practical performance gives us a hint that the existing guarantees are not tight and that there is a need for further investigation of the field. 
\rg{Another direction that should be further explored is the usage of the structure in the signals for designing the sampling operator, as is done for the 2D-DIF operator \cite{Krahmer14Stable, Poon14TV}.}


\section*{Acknowledgment}
The author would like to thank Michael Elad and Yaniv Plan for fruitful discussions.
Raja Giryes is partially supported by AFOSR.
The authors would like to thank the anonymous reviewers for their helpful and constructive comments that greatly contributed to improving 
this paper.


\bibliographystyle{elsarticle-num}
\bibliography{../analysis}

\end{document}